# Electric field control of nonlinear Hall effect in type-II Weyl semimetal TaIrTe$_4$

Jiaju Yang[#], Lujun Wei[#, *], Yanghui Li, Lina Chen, Wei Niu, Shuo Wang, Feng Li,

Ping Liu, Shuang Zhou, and Yong Pu*

*School of Science, Nanjing University of Posts and Telecommunications, Nanjing 210023, P. R. China*

[#] Jiaju Yang and Lujun Wei contributed equally to this work.

*Corresponding authors: wlj@njupt.edu.cn; yongpu@njupt.edu.cn.

## ABSTRACT

The nonlinear Hall effect (NLHE), as an important probe to reveal the symmetry breaking in topological properties of materials, opens up a new dimension for exploring the energy band structure and electron transport mechanism of quantum materials. Current studies mainly focus on the observation of material intrinsic the NLHE or inducing the NLHE response by artificially constructing corrugated/twisted two-dimensional material systems. Notably, the modulation of NLHE signal strength, a core parameter of device performance, has attracted much attention, while theoretical predictions suggest that an applied electric field can achieve the NLHE enhancement through modulation of the Berry curvature dipole (BCD). Here we report effective modulation the magnitude and sign of the NLHE by applying additional constant electric fields of different directions and magnitudes in the type-II semimetal TaIrTe$_4$. The NLHE response strength is enhanced by 168 times compared to the intrinsic one at 4 K when the additional constant electric field of -0.5 kV/cm is applied to the *b*-axis of TaIrTe$_4$ and the through a.c. current is parallel to the TaIrTe$_4$ *a*-axis. Scaling law analysis suggests that the enhancement may be the result of the combined effect of the electric field on the intrinsic BCD and disorder scattering effect of TaIrTe$_4$. This work provides a means to study the properties of TaIrTe$_4$, as well as a valuable reference for the study of novel electronic devices.

**Introduction**

The nonlinear Hall effect (NLHE) is a quantum transport phenomenon in which a transverse second-order voltage response is induced by applying a longitudinal current in the absence of an external magnetic field. Unlike the conventional Hall effect, the generation of the NLHE does not depend on time-reversal symmetry breaking, but requires the system to be characterized by specific low symmetry[1, 2]. The microscopic mechanism are intrinsic and extrinsic contributions: the intrinsic mechanism originates from the geometrical quantum properties of the system, which is manifested as the Berry curvature dipole (BCD) formed by the Berry curvature gradient, which is directly related to the topological nature of the energy bands of the material[2, 3]; and the extrinsic mechanism is mainly related to the side-jump and skew-scattering motions of electrons in disorder systems [4-6]. In theoretical studies, the NLHE has been predicted to possibly exist in a variety of material systems, including two-dimensional (2D) transition mental dichalcogenides[2, 3, 7, 8], the surface of topological crystalline insulators[2], three-dimensional Weyl smimentals[2, 9, 10], and so on. Moreover, theoretical calculations have demonstrated that the generation of the NLHE can be effectively induced by applying lattice strains or constructing interlayer twisted structures[11-15].

Currently, the NLHE have been experimentally discovered in a variety of materials such as $WTe_2$[16-20], $TaIrTe_4$[21], $NbIrTe_4$[22], bulk $BaMnSb_2$[23], $Bi_2Se_3$ surface[24], $Cd_3As_2$[19, 25], $Ce_3Bi_4Pd_3$[26], $T_d$-$MoTe_2$[18, 27], α-(BEDT-TTF)$_2$I$_3$[28] and BiTeBr[29]. Meanwhile, the NLHE has also been successfully observed in materials such as $WSe_2$ and bilayer graphene by interlayer twisting technique[30-34]. In addition, significant progress has been made in experiments utilizing lattice strain for the modulation of the NLHE in WSe$_2$ and WTe$_2$[35, 36]. In particular, a recent theoretical breakthrough has revealed a new mechanism for electric field control of the NLHE: the applied electric field, by driving the Bloch electron displacement operator, can induce the Berry connection polarizativity[34], which then generates the field-excited Berry curvature[37], and its corresponding BCD[38]. This mechanism has been preliminarily experimentally verified in $T_d$-WTe$_2$[38, 39].

As a typical representative of the type-II semimetals, TaIrTe$_4$ exhibits more remarkable the NLHE properties than WTe$_2$: it has been experimentally demonstrated to generate strong NLHE signal at room temperature[21]。In addition to fundamental physical studies, this material shows potential applications in several frontier areas due to its unique electronic structure, such as magnetic storage [40-42] and superconductivity [43, 44], etc. Not only the intrinsic second-order NLHE exists in TaIrTe$_4$[21, 41, 42], but also the third-order NLHE has been observed under specific conditions[45]. More interestingly, the anomalous Hall effect and nonreciprocal Hall effect induced by d.c. current have also been found in thin-layer TaIrTe$_4$[46]. Notably, researchers have recently succeeded in constructing novel devices that enable high-dimensional strain encoding by utilizing their second- and third-order nonlinear Hall voltage signals as independent parameters[36], and this breakthrough advancement has pushed forward the development of NLHE modulation research. However, the NLHE signal strength in device applications is directly related to the sensitivity, and the existing theoretical predictions suggest that the applied electric field may enhance the NLHE signal of the material by modulating the BCD[34, 37, 38, 47, 48]。

In this work, an electric field modulation strategy is proposed based on which effective electric field modulation of the NLHE is realized in TaIrTe$_4$. It is found that when a constant electric field of 0.5 kV/cm is applied along the TaIrTe$_4$ $b$-axis and a a.c. current is applied in the $a$-axis direction, the nonlinear Hall response strength is enhanced by about 168 times compared with the intrinsic state at 4 K. Both the sign and the magnitude of the NLHE can be effectively modulated by the applied electric field. The physical mechanism of the NHLE modulated by electric field is revealed by the scaling law, and the evolution law of BCD with electric field is given quantitatively. This work will provide a valuable reference for the development of high-performance nonlinear Hall devices.

**Results**

**Structure, resistance and Raman spectra for the TaTrTe$_4$ device**

The surface of the TaIrTe$_4$ crystal structure has a clear uniaxial mirror symmetry with mirror lines ($M_a$) along the lattice $b$-axis and no mirror symmetry along the $a$-axis, as shown in Figure 1(a). Figure 1(b) shows a photomicrograph of the device A1 structure, which consists of a mechanically exfoliated TaIrTe$_4$ flake transferred onto a disk-shaped electrode. As a special note, all core data in the main paper are extracted from this device. The thickness of the TaIrTe$_4$ flake of the device was measured by atomic force microscopy (AFM) to be about 88.2 nm (see Figure S1 of supplementary material). The NLHE was investigated by an in-plane d.c. electric field ($\boldsymbol{E}_{dc}$) tuning strategy, and the circuit configuration is shown in Figure 1(c). The resistive anisotropy of the device is shown in Figure 1(d), indicating that it is highly correlated with the crystal structure, with the resistance of its $b$-axis approximately 4.4 times higher than that of its $a$-axis, which is consistent with the reported previously[21]. Raman spectra were collected by rotating the incident laser polarization at different angles with respect to the $a$-axis of TaIrTe$_4$ and the integrated intensity under each peak was calculated and plotted as contour plots as shown in Figure 1(e). Figure 1(f) shows the dependence of the Raman peak intensity on the polarization angle at 147.4 cm$^{-1}$, which is in agreement with the previously reported results[41, 49]. The peak intensity maxima are located at the $a$-axis[49], which indicates that the 0° and 90° directions correspond to the $a$-axis and $b$-axis, respectively, of the TaIrTe$_4$ flake.

**Electric field control of NLHE**

An a.c. electric field $E^\omega$ was applied in the $a$-axis direction of TaIrTe$_4$ while $\boldsymbol{E}_{\text{dc}}$ was applied along the $b$-axis (i.e., θ = 90°). When $E_{\text{dc}}$ = 0 kV/cm, a very weak second-order anomalous Hall voltage $V_H^{2\omega}$ was measured in the $b$-axis direction (Figure 2(a)). However, the $V_H^{2\omega}$ was significantly enhanced when $E_{\text{dc}}$ = ±0.3 or ±0.5 kV/cm, as shown in Figure 2(a). We have excluded external effects such as diode effect, thermal effect and thermoelectric effect (see Figures S2-S6 of supplementary material) that may cause $V_H^{2\omega}$, and further confirmed that $V_H^{2\omega}$ is induced by an applied constant electric field. When the constant electric field is positive ($\boldsymbol{E}_{\text{dc}} > 0$), the measured $V_H^{2\omega}$ signal is negative and increases with increasing electric field strength. On the contrary, when $\boldsymbol{E}_{\text{dc}} < 0$, the measured $V_H^{2\omega}$ signal is positive and increasing electric field strength. When $\boldsymbol{E}_{\text{dc}}$//$a$-axis (i.e., θ = 0°), the $V_H^{2\omega}$ signal weakens, as shown in Figure 2(b). Similarly, when $E^\omega$//$b$-axis, $E_{\text{dc}}$//$a$-axis, a very significant second-order anomalous Hall voltage is observed in the $a$-axis direction, whereas the second-order signal is weakened or even disappeared at $E_{\text{dc}}$//$b$-axis as shown in Figures 2(c) and 2(d), which is similar to the phenomenon reported in the Wely semimetal WTe$_2$[38].

This phenomenon can be initially explained by the fact that when an external constant electric field is applied on the device, a larger BCD, whose magnitude was denoted as ***D***, is induced along the direction parallel to the alternating electric field. Since the second-harmonic Hall voltage is proportional to ***D***·***E***$^\omega$[2], after inputing an small a.c. current is applied along the *a*-axis or *b*-axis, strong $V_H^{2\omega}$ signals are detected along the *b*-axis or *a*-axis when ***E***$_{dc}$⊥***E***$^\omega$ (Figures 2(a) and 2(c)). However, the $V_H^{2\omega}$ signal weakens or disappearars when ***E***$_{dc}$∥***E***$^\omega$ (Figures 2(b) and 2(d)). In addition, similar results are observed in devices A2, A3 and A4 with varying thicknesses (see Figures S8, S10 and S12 of supplementary material). In the present work, the thicknesses of the TaIrTe$_4$ flakes used are greater than 60 nm, and the second-harmonic Hall voltage signals are measured to be very wery weak at room temperature or low temperatures without an external constant electric field (see Figure S13 of supplementary material), which is agreement with the reported previously[21].

Next, we investigated the dependence properties of the electric field-controlled NLHE with ***E***$_{dc}$ angle. The measured $V_H^{2\omega}$-$I^\omega$ curves when applying ***E***$_{dc}$ = 0.5 kV/cm at different angles are shown in Figures 3(a) and 3(b). It can be observed that the $V_H^{2\omega}$-$I^\omega$ curves change significantly when ***E***$_{dc}$ is applied at different angles. To better present its angle-dependent characteristics, we calculated the second-order nonlinear Hall response strength $[E_H^{2\omega}/(E^\omega)^2]$ at each angle and plot its angle-dependent curves as shown in Figures 3(c) and 3(d). $E_H^{2\omega}$ and $E^\omega$ can be obtained by the formulas $E_H^{2\omega} = V_H^{2\omega}/W$ and $E^\omega = I^\omega R/L$, where W, R and L are the channel width, the longitudinal resistance and the channel length, respectively. The value of $[E_H^{2\omega}/(E^\omega)^2]$ is the slope obtained by linearly fitting the data of $E_H^{2\omega}$ to $(E^\omega)^2$ (see Figure S14 of supplementary material). When $E_{dc}$ was not applied, the second-order Hall response signal was very weak, which is consistent with the previous results. When a constant $E_{dc}$ is applied, the second-order nonlinear Hall response srtrength $[E_H^{2\omega}/(E^\omega)^2]$ shows significant anisotropy (Figures 3(c) and 3(d)), which is highly correlated with the structural characteristics of TaIrTe$_4$.

When $E^\omega$//a-axis and $E_{dc}$ = 0.5 kV/cm, the value of $[E_H^{2\omega}/(E^\omega)^2]$ immediately changes from positive to negative and gradually decreases with the increase of $\theta$ until it reaches a minimum at $\theta$ = 90°, and then it gradually increases as $\theta$ increases to 180°. When $\theta$ increases from 180° to 360°, the change of $[E_H^{2\omega}/(E^\omega)^2]$ is opposite to its increase from 0° to 180° (Figure 3(c)). The variation of $[E_H^{2\omega}/(E^\omega)^2]$ and $\theta$ when a reversed external electric field is applied is opposite to that when a positive electric field is applied. The maximum $[E_H^{2\omega}/(E^\omega)^2]$ = 1.72×10$^{-4}$ m/V is observed when $E^\omega$//$a$-axis, $E_{dc}$ = -0.5 kV/cm, and $\theta$ = 90°, which is increased by a 168 times as compared with that when no external field is applied (1.02×10$^{-6}$ m/V ). At any $\theta$, $[E_H^{2\omega}/(E^\omega)^2]$ increases with $E_{dc}$, indicating a positive correlation with $E_{dc}$. The electric field-induced control of $[E_H^{2\omega}/(E^\omega)^2]$ (Figure 3(c)) at $\theta \neq$ 90°/270° may be due to the BCD component $D_a$ generated along the $a$-axis and $D_a$//$E^\omega$. Meanwhile, we also observed that the $V_H^{2\omega}$ signal was measured at $\theta$ = 0° and $E_{dc} \neq 0$ when $E^\omega$//a-axis (Figures 2(b) and 3(a)), which may be due to the fact that the transferred TaIrTe$_4$ $a$-axis is not perfectly parallel to the electrodes at $\theta$ = 0°, which results in the generation of a smaller $D_a$. At $E^\omega$//$b$-axis, the opposite angular dependence law as at $E^\omega$//$a$-axis was observed (Figures 3(d)). In addition, similar results were obtained for devices A2 and A3 (see Figures S8 and S10 of supplementary material), which indicate the reliability of the results.

**Mechanism of NLHE controlled by electric field**

In order to further clarify the mechanism of electric field control of NLHE, we investigated the temperature-dependent properties and scaling law of the second-order NLHE. Figures 4(a) and 4(b) show the results of the $V_H^{2\omega}$ signals as a function of temperature at $\theta$ = 90° and 0°, respectively. The $V_H^{2\omega}$ signal decreases with increasing temperature for both $E^\omega$//$a$-axis or $E^\omega$//$b$-axis, which may be due to the thermal perturbation caused by increasing temperature masking the NLHE signal[35]. Notably, the device is still measured with a significant $V_H^{2\omega}$ signal when $E_{dc}$ = 0.5 kV/cm is applied at room temperature relative to the no external field case (see Figure S15 of supplementary material), suggesting that the electric-field-induced NLHE in TaIrTe$_4$ can be maintained up to room temperature. We observed similar results in device A4 (see Figure S12 of supplementary material).

Figures 4(c) and 4(d) give the relationship curves between the second-order NLHE response strength $[E_H^{2\omega}/(E^\omega)^2]$ and $\sigma_{xx}$ under the application of different external fields, where $\sigma_{xx}$ is the longitudinal conductivity at different temperatures, which can be obtained by calculating $\sigma_{xx}$ = (1/$R$)($L$/$Wd$), where $d$, $L$, and $W$ are the thickness, length, and width of the TaIrTe$_4$ flake, respectively. According to Fu et al.[2], the scaling law between $[E_H^{2\omega}/(E^\omega)^2]$ and $\sigma_{xx}$ satisfies the equation $[E_H^{2\omega}/(E^\omega)^2]$ =

$C_2\sigma_{xx}^2+C_1\sigma_{xx}+C_0$, where $C_0$, $C_1$, and $C_2$ are constants. $C_2$ and $C_1$ involve the contribution of external factors such as side-jump, and $C_0$ is mainly contributed by intrinsic BCD.

As shown in Figures 4(c) and 4(d), the scaling law formulation fits the results of the external electric field modulation well. By analyzing the variation of the fitted parameter values (Tables S1 and S2 of supplementary material), we consider that the electric-field-induced NLHE in TaIrTe$_4$ may be a result of the combined effect of BCD and disordered scattering contributions. We also estimated the value of the electric field-induced BCD, and the value of the BCD excited by the external field can be calculated by $\boldsymbol{D} = (2\hbar n/m^*e)[E_H^{2\omega}/(E^\omega)^2]$, where $\hbar$, $m^*$, $n$, and $e$ are the approximate Planck's constant, the effective electron mass, the carrier concentration, and the electron charge, respectively. Replacing $[E_H^{2\omega}/(E^\omega)^2]$ with the coefficient $C_0$ obtained from the scaling law fit[38], we calculated the magnitude of the BCD at $\theta = 90°$ and $\boldsymbol{E^\omega}//a$-axis, and $\theta = 0°$ and $\boldsymbol{E^\omega}//b$-axis, and denote them as $\boldsymbol{D}_a$ and $\boldsymbol{D}_b$, respectively. The estimated $\boldsymbol{D}_a$ and $\boldsymbol{D}_b$ at different $\boldsymbol{E}_{dc}$ are shown in Figures 4(e) and 4(f), respectively, and these two quantities exhibit a clear change of direction when switching the direction of the electric field and the magnitudes show a monotonic relationship with the electric field.

In addition, we analyzed the variation of the field-excited BCD with the angle of the electric field at $\boldsymbol{E}_{dc} = 0.5$ kV/cm (see Figure S16 of supplementary material), which indicates that the regulation of the BCD magnitude and orientation can be achieved by changing both the magnitude and the direction of the electric field.

**Conclusions**

In this work, we have observed significant modulation of NLHE using an applied direct current electric field in the type-II Weyl semimetal TaIrTe$_4$. By varying the magnetude and direction of the electric field, we were able to effectively modulate the magnitude and sign of NLHE. Scaling law analysis indicates that the mechanism of the electric field control of NLHE arises from the combined effects of the intrinsic BCD and external disordered scattering contributions. Therefore, the enhancement of the electric field-tuned NLHE in TaIrTe$_4$ greatly increases its potential for applications, providing a promising candidate for the development of novel electronic devices. This finding not only offers new insights into the physical mechanisms of NLHE but also paves the way for future electronic device design.

**Experimental Section**

Sample Preparation: First, flakes were obtained from bulk single crystal $T_d$-TaIrTe$_4$ (HQ Graphene) by mechanical exfoliation. Then, disk-shaped Ti(5 nm)/Au(20 nm) was prepared on a Si/SiO$_2$ (285 nm) substrate using e-beam lithography and e-beam evaporation techniques. Finally, Td-TaIrTe$_4$ flakes were transferred to this electrode by dry transfer. To avoid oxidation, both exfoliation and transferring of the flakes were performed in a glove box filled with N$_2$ (water, oxygen content <0.1 ppm).

Sample Characterization: The thickness of TaIrTe$_4$ flakes was measured using AFM (Dimension ICON). A Raman spectroscopy instrument (Renishaw inVia) with a laser wavelength of 532 nm was used to characterize the crystal structure of TaIrTe$_4$ flakes. Direct current and sinusoidal a.c. current with a frequency of 13.7 Hz were applied through sourcemeters Keithley 2450 and 6221, respectively, while the first/second harmonic test signals were performed with a lock-in amplifier (Stanford 830).

cknowledgement

This work was supported by the National Natural Science Foundation of China (Nos. 52001169, and 61874060), the Open Foundation from Anhui key laboratory of low-energy quantum materials and devices, and the Innovation Project of Jiangsu Province (No. KYCX22_0918).

Conflict of Interest

The authors have no conflicts to disclose.

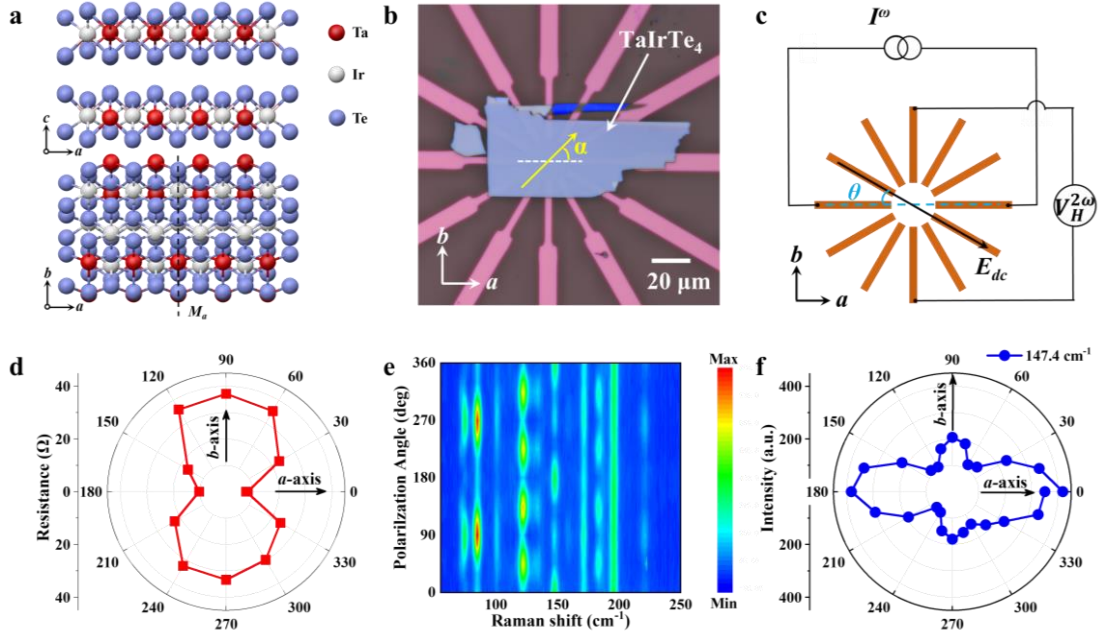

**Figure 1.** (a) TaIrTe$_4$ crystal structure. (b) Device A1 optical picture. (c) Measurement configuration for the NLHE with $E^{\omega}$//$a$-axis. $\theta$ is defined as the angle between $E_{dc}$ and $a$-axis. (d) Resistance of the TaIrTe$_4$ flake as a function of $\theta$. (e) Polarization Raman spectrum of the TaIrTe$_4$ flake. $\alpha$ is defined as the polarization angle in (b). (f) Dependence of Raman spectrum intensity at 147.4 cm$^{-1}$ with $\alpha$.

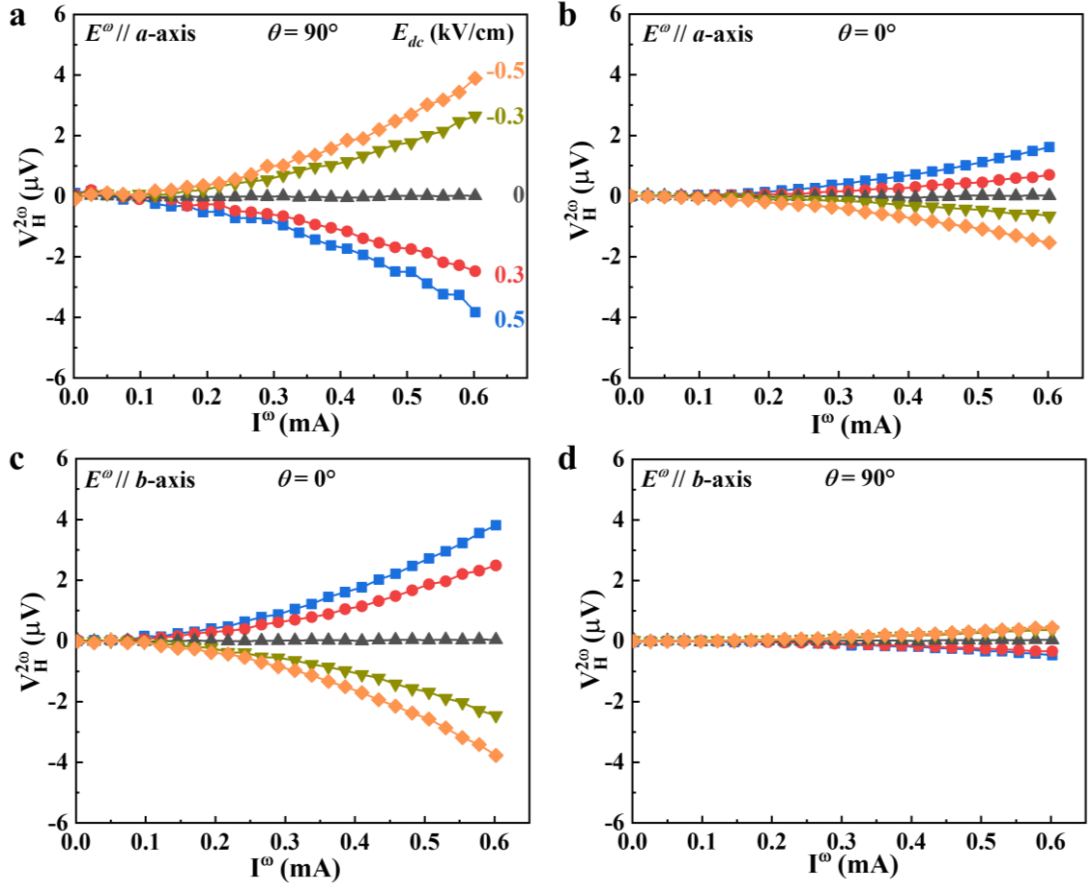

**Figure 2.** The second-order harmonic Hall voltage $V_H^{2\omega}$ versus $I^\omega$ measured by $E^\omega$//$a$-axis and at 4 K under different $E_{dc}$ with (a) $E_{dc}$//$b$-axis ($\theta = 90°$) and (b) $E_{dc}$//$a$-axis ($\theta = 0°$), respectively. $V_H^{2\omega}$ versus $I^\omega$ measured by $E^\omega$//$b$-axis and at 4 K under different $E_{dc}$ with (a) $E_{dc}$//$a$-axis ($\theta = 0°$) and (b) $E_{dc}$//$b$-axis ($\theta = 90°$), respectively.

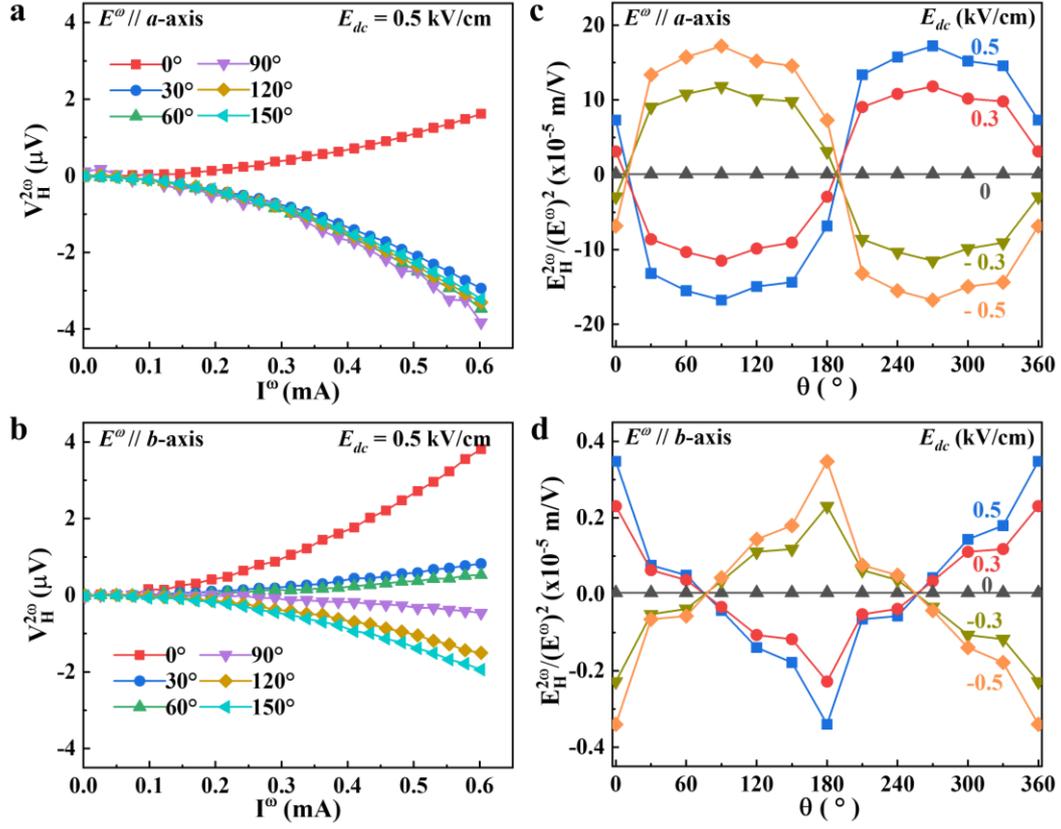

**Figure 3.** $V_H^{2\omega}$ versus $I^\omega$ curves measured at 4 K and $E_{dc}$ = 0.5 kV/cm along different angles with (a) $E^\omega$//$a$-axis and (b) $E^\omega$//$b$-axis, respectively. $[E_H^{2\omega}/(E^\omega)^2]$ versus $\theta$ with (c) $E^\omega$//$a$-axis, and (d) $E^\omega$//$b$-axis at 4 K..

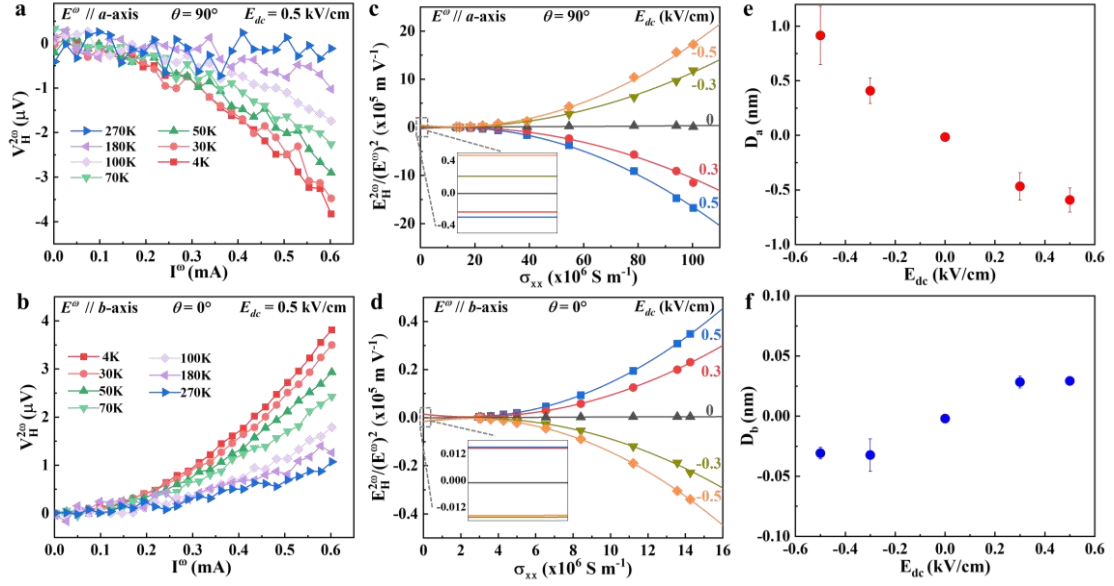

**Figure 4.** $V_H^{2\omega}$ versus $I^\omega$ with (a) $E^\omega$//a-axis, and (b) $E^\omega$//b-axis at different temperatures and $E_{dc}$ = 0.5 kV/cm. $[E_H^{2\omega}/(E^\omega)^2]$ versus $\sigma_{xx}$ with (c) $E^\omega$//a-axis and $\theta$ = 90°, and (d) $E^\omega$//b-axis and $\theta$ = 0°. The solid line is the result of fitting with the scaling law equation. The inset shows an enlarged view at the intersection of the fitted curve and the vertical axis. (e) $D_a$ and (f) $D_b$ at different $E_{dc}$.